\documentclass[journal=jctcce,manuscript=article,layout=twocolumn]{achemso}

\usepackage{amsmath} 
\usepackage{siunitx} 
\usepackage{graphicx} 
\usepackage[shortcuts]{extdash} 
\usepackage{hyperref} 

\author{Indranil Mal}
\email{mal@fzu.cz}
\affiliation{FZU - Institute of Physics of the Czech Academy of
  Sciences\\ Na Slovance 2, Prague, 18 200, Czech Republic}
  
\author{Milan Ko\v{c}\'{i}}
\affiliation{FZU - Institute of Physics of the Czech Academy of
  Sciences\\ Na Slovance 2, Prague, 18 200, Czech Republic}
\alsoaffiliation{Faculty of Nuclear Sciences and Physical Engineering,
  Czech Technical University in Prague, B\v{r}ehov\'{a} 7, Prague, 115 19, Czech Republic}

\author{Paolo Nicolini}
\affiliation{FZU - Institute of Physics of the Czech Academy of
  Sciences\\ Na Slovance 2, Prague, 18 200, Czech Republic}

\author{Prokop Hapala}
\email{hapala@fzu.cz}
\affiliation{FZU - Institute of Physics of the Czech Academy of
  Sciences\\ Na Slovance 2, Prague, 18 200, Czech Republic}

\title{GridFF: Efficient Simulation of Organic Molecules on Rigid Substrates}


\begin{document}

\begin{tocentry}
  \includegraphics[width=3.25in,height=1.75in]{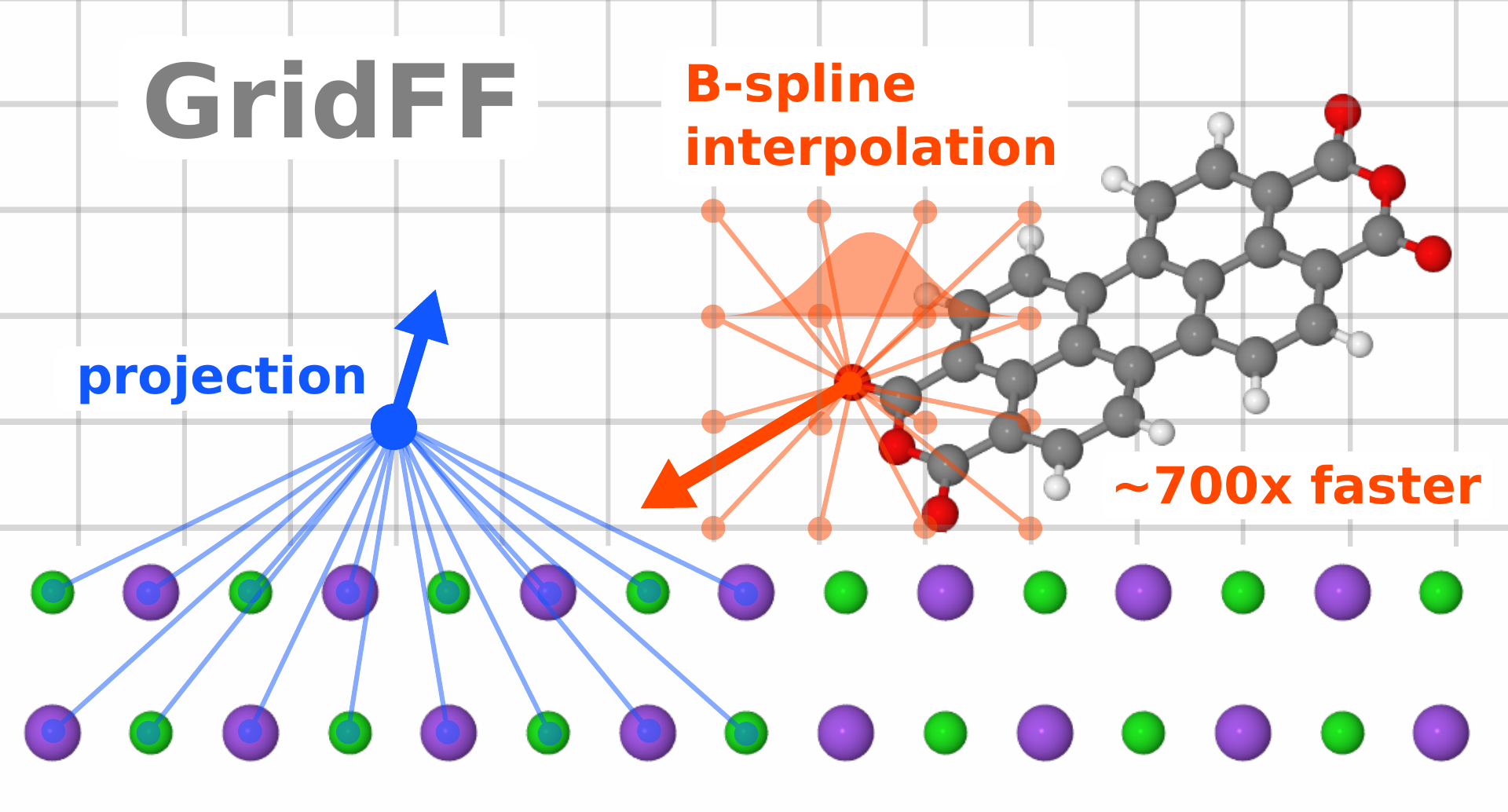}
  For Table of Contents Only
\end{tocentry}

\begin{abstract}
  We present GridFF, an efficient method for simulating molecules on
  rigid substrates, derived from techniques used in protein\-/ligand
  docking in biochemistry.
  By projecting molecule\-/substrate interactions onto precomputed
  spatial grids with tricubic B-spline interpolation, GridFF reduces
  the computational cost by orders of magnitude compared to
  traditional pairwise atomistic models, without compromising the
  accuracy of forces or trajectories.
  The CPU implementation of GridFF in the open\-/source FireCore
  package provides a 100-1000$\times$ speedup over all-atom simulations using
  LAMMPS, while the GPU implementation -- running thousands of system
  replicas in parallel -- samples millions of configurations per
  second, enabling an exhaustive exploration of the configuration
  space of small flexible molecules on surfaces within minutes.
  Furthermore, as demonstrated in our previous application of a
  similar technique to high-resolution scanning probe microscopy,
  GridFF can be extended beyond empirical pairwise potentials to those
  derived from \emph{ab initio} electron densities.
  Altogether, this unlocks accurate high-throughput modeling of
  molecular self-assembly, adsorption, and scanning probe manipulation
  in surface science.
\end{abstract}

\section{Introduction}\label{sec:intro}

The structural characterization and design of organic/inorganic
interfaces represent a critical challenge across multiple research and
technological fields, including friction and lubrication\cite{Krim12,
  Villa24, Kawai16}, molecular electronics\cite{Chen21},
photovoltaics\cite{Surin06}, and scanning probe microscopy
(SPM)\cite{Hieulle18}. The interaction of organic molecules with
surfaces of inorganic crystals plays a key role in the development of
molecular nanotechnology comprising self\-/assembled
monolayers\cite{Heimel08}, 2D molecular crystals\cite{Feng21},
covalent organic frameworks\cite{Zhu24}, and also for the emerging
field of on\-/surface chemistry\cite{George09}. Complex molecular
nanostructures are one of the most promising building blocks for
next\-/generation computational devices such as molecular
electronics\cite{Yuan24}, photonics\cite{Ying21}, and quantum cellular
automata\cite{Chen22}.

The formation of such structures is primarily governed by
non\-/covalent interactions between organic molecules and the
templating effects of the inorganic substrate. However, predicting the
absorption configurations of molecules on surface and self\-/assembled
layers remains an open challenge, akin crystal structure prediction in
pharmaceuticals\cite{Beran23} and protein folding in
biochemistry\cite{Nassar21}. The challenge of designing such
structures, especially from flexible molecules, arises from the curse
of dimensionality, related to soft internal degrees of freedom
(\emph{e.g.,} torsions along single bonds), which drastically expands
their configuration space. Fortunately, the interaction with rigid
inorganic substrates can constrain this structural freedom, enabling
more systematic and rational design of self\-/assembled motifs from
first principles.

Crystalline inorganic substrates also provide an ideal, well\-/defined
support for the experimental construction and study of complex
nanosystems\cite{Li24}. Unlike bulk solids or liquid solutions,
surfaces allow unobstructed access to functional molecular components
and atomic\-/level control through SPM techniques. SPM is not only an
indispensable tool for imaging nanostructures with atomic
resolution\cite{Sumaiya23}, but it also offers the potential to
construct complex supramolecular structures by direct
manipulation\cite{Xiong23}. However, the optimal control of molecular
degrees of freedom through interactions with an atomic force
microscope (AFM) or a scanning tunneling microscope (STM) tip remains
a major challenge, heavily dependent on atomistic simulations. Without
such simulations, SPM manipulation becomes a blind trial\-/and\-/error
process, as the instrument provides minimal experimental information
about molecular configurations during
manipulation\cite{Scheidt23}. Recent efforts to automate this
laborious process involve reinforcement learning for robotic AFM/STM
machines\cite{Leinen20, Ramsauer23, Ramsauer24}. However, training
such robotic systems through direct experimentation is prohibitively
expensive and time\-/consuming due to the vast number of required
training examples. This underscores the need for a specialized virtual
training nanophysics engine -- akin to NVIDIA’s Isaac physics
engine\cite{Liang18} -- capable of generating the data needed for
training robotic manipulation of nanoscale objects, possibly
parallelized on graphics processing units (GPU).

An even more ambitious challenge is the computational molecular design
of organic molecules that deterministically self\-/assemble into
target patterns or shapes on a substrate, \emph{e.g.} driven by
hydrogen bond formation\cite{Manikandan24}. In this case, whilst
chemical intuition may suffice for designing simple structures based
on one or two hydrogen bonds from rigid molecules, creating complex
patterns requires molecules with significantly greater structural
information and an inverse computational design. Notable examples
include biopolymers such as DNA, RNA, and proteins, whose
self\-/assembly can be computationally designed using automated tools,
often leveraging machine learning approaches like
AlphaFold\cite{Jumper21}. However, training such design tools requires
extensive databases of self\-/assembled (folded) structures, which in
biochemistry are obtained from experimental data or atomistic
simulations\cite{PDB18}. In the domain of molecular surface science,
no comparable database currently exists, and constructing such a
database for newly designed molecules using experimental methods would
be prohibitively expensive. Therefore, softwares capable of
efficiently predicting self\-/assembled patterns for a broad range of
novel molecules are essential for the rational design of molecular
nanostructures on surfaces.

Moreover, the structure of molecular assemblies at finite temperatures
is influenced not only by enthalpy, but also by entropic
contributions. For soft, flexible molecules, entropy can play a
particularly significant role. The temperature dependence of the
entropy term in the Gibbs free energy ($\Delta G = \Delta H - T \Delta
S$) is crucial for estimating the melting temperature of
self\-/assembled structures and determining annealing conditions for
their reversible formation. These are key parameters in computational
design, particularly for larger self\-/assembling molecular templates
akin to DNA fragments or oligopeptides. However, the computational
estimation of entropy and free energy requires an exhaustive
configuration sampling of the partition function, which is extremely
demanding -- especially when one aims at screening a large number of
candidate molecules.

Consequently, both SPM manipulation of molecules on solid substrates
and their self\-/assembly ultimately face the same fundamental
challenge: exploring a vast configuration space of soft organic
molecules while simultaneously describing their interactions with the
substrate (and potentially with an AFM/STM tip) efficiently. The sheer
size of this configuration space renders \emph{ab initio} methods
computationally unfeasible, necessitating the use of classical force
fields. Unfortunately, many existing methods for molecular
configuration exploration, such as CREST\cite{Pracht24} -- designed
for systems in gas or liquid phases -- cannot be directly applied to
molecules on surfaces. Likewise, state\-/of\-/the\-/art force fields,
like AMBER\cite{Wang04}, CHARMM\cite{Vanommeslaeghe10},
GROMOS\cite{Oostenbrink04} and OPLS\cite{Jorgensen96} (which can be
used on general\-/purpose simulation programs such as
AMBER\cite{Case05}, CHARMM\cite{Hwang24}, GROMACS\cite{Abraham15},
LAMMPS\cite{Thompson22} and NAMD\cite{Phillips20}), are primarily
optimized for biological molecules in aqueous environments, making
efficient simulations of molecular interactions with substrates highly
challenging.

To address these limitations and provide an efficient tool for SPM
manipulation and self\-/assembly of small organic molecules on ionic
substrates, we have developed a new classical force field simulation
software\cite{FireCore}. The software we introduce here is optimized
for efficient parallel configuration sampling, accelerated using
modern GPUs, and incorporates a specialized grid\-/projected force
field (GridFF) to enhance the description of molecule\-/substrate
interactions. This approach, which is the main focus of this
publication, enables precise and computationally feasible predictions
of molecular self\-/assembly and manipulation on inorganic surfaces,
bridging the gap between theoretical modeling and experimental
nanostructure design.

\section{Motivations}\label{sec:motivation}

Simulating organic molecules on inorganic substrates, even using
classical force fields, can be relatively costly, particularly when
using large supercells comprising a large number of atoms. In such
systems the number of substrate atoms $n_S$, replicated across
multiple unit cells, far exceeds the number of atoms of the organic
molecules $n_M$ being studied, leading to substantial computational
overhead. There is more than one reason why one has to use large
supercells. First of all, in order to properly model a crystal slab,
several atomic layers shall be considered. In complex surface
reconstructions, the unit cell itself may be also very
large. Moreover, if one aims to simulate an isolated molecule on a
surface, the sizes of the cell must be large enough to avoid artifacts
due to the interaction between the molecule and its periodic
images. On the other hand, even in the case of periodic systems such
as self\-/assembled monolayers, one has to deal with the intrinsic
incommensurability between the two lattices. For all these reasons,
the typical case is $n_S \gg n_M$. Furthermore, many classical force
fields parameterizations struggle to simultaneously capture the
nuances of both molecule\-/substrate interactions, and bulk mechanics
of the crystal. For this reason, in such simulations, it is a common
practice to fix the position of the substrate atoms, or at least the
bottom atomic layers. Even in this case, the computational overhead is
not reduced in most implementations as the number of pairwise
interaction scales quadratically with the total number of atoms in the
supercell $(n_M+n_S)^2$. Even if pairwise interactions between the
atoms of the substrate can be eliminated (which is generally a
non\-/standard feature in classical force field
programs\cite{note_lammps}), the computational cost of evaluating $n_M
n_S$ pairwise interactions between the molecule and the substrate
dominates over the $n_M^2$ calculations required to describe the
interactions between atoms in the molecule.

Electrostatics interactions, due to their slow convergence in
real\-/space calculations, require a different treatment. They are
usually addressed by evaluating the interactions in the Fourier space
using techniques derived from the Ewald summation\cite{Ewald21}, such
as the Particle\-/Mesh Ewald (PME) method\cite{Darden93}. However,
achieving a satisfactory accuracy requires a large plane\-/wave cutoff
and a correspondingly dense grid, making PME a frequent bottleneck in
periodic classical force field calculations. This computational cost
increases with the supercell size as $O(N \log(N))$ (where $N$ is the
number of grid points).

GridFF, inspired by a similar method used typically for rigid
protein\-/ligand docking\cite{Pattabiraman85, Meng92}, offers an
alternative approach designed to address these limitations. Unlike
traditional force fields, which evaluate all pairwise interactions and
PME on\-/the\-/fly, GridFF precalculates the molecule\-/substrate
interaction potential on a grid before the actual simulation is
performed. This precalculated grid, representing the interaction
potential, is then simply interpolated during the simulation. This
largely decreases the computational cost, and allows one to use denser
grids, larger supercells, and more complex force fields without added
computational cost. In addition to this, GridFF does not necessarily
need to be constructed from pairwise potentials (\emph{e.g.}, as in
the D3 van der Waals corrections\cite{Grimme10}, where three\-/body
terms are considered), but can also be computed directly from electron
densities obtained from \emph{ab initio} wavefunctions, as it is
routinely done in the so\-/called full density based model (FDBM) in
atomic force microscopy simulations\cite{Ellner19, Oinonen2024}.

Whilst the computational cost of a traditional pairwise potential
scales quadratically with the number of particles in the system, and
Ewald summation scales logarithmically (albeit with a significant
prefactor), the cost of evaluating the GridFF potential through
interpolation remains essentially constant. Although cache memory
effects might subtly influence the performance for large grids, the
computational complexity of the interpolation algorithm itself is
independent of the grid size, offering significant potential for
speedup when the molecule\-/substrate interaction is the computational
bottleneck.

\section{Principles}\label{sec:principles}

Grid\-/based potentials for the description of non\-/covalent
interactions, such as Coulomb and Lennard\-/Jones, are routinely used
in computational biochemistry codes for protein\-/ligand
docking\cite{Friesner04, Allen15, Eberhardt21}. Nevertheless, here we
briefly recall the approach originally published by
Goodford\cite{Goodford85}. The core idea lies in the assumption that
the total interaction energy $E$ between the molecule and the
substrate can be decomposed into a sum of atomic contributions
\begin{equation}
  E = \sum_i E_i(r_i,t_i,Q_i),
  \label{eq:E}
\end{equation}
where $r_i$, $t_i$ and $Q_i$ are the position, type and charge of the
molecular atom $i$. Each term $E_i$ can be interpolated from the grid
potential using efficient techniques such as 3D
trilinear\cite{Diller99} or tricubic splines\cite{Oberlin98}, which
are particularly well\-/suited for implementation on GPU hardware. In
ligand docking applications, the trilinear approximation prevails due
to the minimal computational cost, despite the fact that it cannot
provide high quality forces\cite{Minh18}. Grid potentials are then
typically used to evaluate energy\-/based scoring function only, not
to run molecular dynamics simulations or force\-/based
optimizations\cite{Tomioka94}. Moreover, these potentials are often
artificially softened to effectively mimic flexibility and thermal
fluctuations of the protein at room
temperature\cite{Venkatachalam03}. In some approaches, the grid force
fields are used for mapping free energy profiles using Monte Carlo
sampling at finite temperature\cite{Forouzesh17, Minh20, Ren23}. In
such situations, eventual inaccuracies in energy and force evaluation
are smeared out by thermal fluctuations.  This is in contrast to our
applications of GridFF for force\-/based dynamical simulations at low
temperature, where molecule precisely follows the potential energy
surface (PES), and a delicate balance between the gradient of the PES
and the driving force (from the AFM tip) determines bifurcations of
the trajectory near saddle points. Therefore, in this work we opted
for higher order interpolation (\emph{i.e.}, tricubic B\-/splines),
and much finer grids ($\sim$0.1 \AA) to obtain high\-/quality
forces. We found that even this computationally more intensive
settings for GridFF provide significant speedups compared to
all\-/atom potentials.

As already mentioned, a grid\-/based potential can be constructed
using any desired method, namely a direct projection of pairwise
potentials, a fitting procedure based on density functional theory
(DFT) calculations, or a machine\-/learned potential. In its simplest
form, GridFF approximates the molecule\-/substrate interaction by
projecting typical non\-/covalent pairwise potentials -- such as
Coulomb, Morse, or Lennard\-/Jones -- onto the grid. In the general
case, this involves evaluating the potential
$$E_i(r_i,t_i,Q_i) = \sum_j V_{ij}(|r_i-r_j|, t_i,t_j, Q_i,Q_j)$$
for each atomic type $t_i$ of the molecular atoms interacting with all
substrate atoms $j$. In principle this is doable; nevertheless, it can
be very memory demanding, as one may end up with tens or hundreds of
grids $V_{ij}$ for all possible $ij$ combinations, each accounting
hundreds of megabytes of memory.

Another consideration to take into account is the efficiency of cache
memory (which decreases when reading from distant memory addresses),
finally making this formulation practically useless. For this reason,
grid\-/based potentials often\cite{Pattabiraman85, Meng92,
  Venkatachalam03} (but not always\cite{Oberlin98, Diller99, Wu03})
rely on various factorization schemes where the parameters of the
ligand atoms (the molecular adsorbate in our case) can be factored out
of the summation over the substrate atoms.  This can be done easily
for Coulomb and Morse potential with Lorentz\-/Berthelot mixing
rules. For electrostatics interactions, this is trivial:
\begin{align}
  E_i^{C}(r_i,Q_i) & = \sum_j \frac{Q_i Q_j}{|r_i - r_j|} = Q_i
  \sum_j \frac{Q_j}{|r_i - r_j|} \nonumber \\
  & = Q_i V_i^{C}(r_i).
  \label{eq:E_Coulomb}
\end{align}
The factorization for a Morse potential can be done by applying basic
properties of the exponential function (\emph{i.e.}, $e^{a+b} = e^a
\cdot e^b$) separately to the repulsive (Pauli) and attractive (London
dispersion) parts:
\begin{align}
  \label{eq:E_Morse}
  E_i^{M}(r_i,t_i) = &
  \sum_j \varepsilon_i \varepsilon_j e^{-2 \alpha \left( |r_i - r_j| - R_i - R_j \right)} \nonumber \\
    & - 2 \sum_j \varepsilon_i \varepsilon_j 
    e^{-\alpha \left( |r_i - r_j| - R_i - R_j \right)} 
   \\
  = & P_i(t_i) V_i^{P}(r_i) + L_i(t_i) V_i^{L}(r_i) \nonumber ,
\end{align}
where 
\begin{align}
  P_i(t_i) &= \varepsilon_i e^{2 \alpha R_i} \nonumber \\
  L_i(t_i) &= -2 \varepsilon_i e^{\alpha R_i} \nonumber \\
  V_i^{P}(r_i) &= \sum_j \varepsilon_j e^{-2 \alpha \left(|r_i -
    r_j| - R_j \right)} \nonumber \\
  V_i^{L}(r_i) &= \sum_j \varepsilon_j e^{-\alpha \left(|r_i -
      r_j| - R_j \right)}.
  \label{eq:V_Morse}
\end{align}
Therefore, the three grids defined above ($V_i^{C}$,
$V_i^{P}$ and $V_i^{L}$) do not carry any dependence on
molecular atomic types nor charges. This means that once the grids
have been calculated, then it is possible to simulate the interaction
with any molecule, without the need of any other precomputation.
It is also worth noting that the GridFF approach does not pose any
restriction on the number of substrate atom types.  Finally, during
the evaluation of energies and forces, we simply perform the
interpolation of the three separately stored grids (with the
coefficients in Eq. \ref{eq:V_Morse}) using Eqs. \ref{eq:E},
\ref{eq:E_Coulomb} and \ref{eq:E_Morse}, and considering that $E =
\sum_i E_i^{C} + E_i^{M}$.

\section{Periodicity and long-range electrostatics}\label{sec:ewald} 

As anticipated in the previous section, the treatment of long\-/range
Coulomb interactions in classical force field simulations requires
some attention. A naive evaluation of electrostatic energy via direct
pairwise summation does not converge for periodic systems, and even in
neutral systems, electrostatic forces converge only very slowly with
distance, consuming a relevant amount of computational
resources. Fortunately, periodic boundary conditions (PBC) allow the
use of efficient reciprocal\-/space summation techniques based on the
Ewald summation. For example in PME\cite{Darden93}, atomic charges are
projected onto a grid as a charge density $\rho(\vec{r})$, and the
electrostatic potential is calculated in the reciprocal space via
$V(\vec{k}) = \rho(\vec{k})/|\vec{k}|^2$, using fast Fourier
transforms (FFT) to switch between real and reciprocal space
representations.

The PME algorithm is so efficient that it is often used even for
systems that are not naturally periodic (\emph{e.g.}, proteins in
water, or a single molecule on a surface), at the cost of introducing
artificial interactions between periodic images of the system. Using
PME for surfaces is further complicated by the lack of periodicity in
the $z$\-/direction. A well\-/established and computationally
efficient solution (used both in LAMMPS and FireCore) is to add a
sufficient vacuum padding above the surface and apply monopole and
dipole corrections to the resulting 3D\-/periodic solution, assuming
that higher\-/order multipole contributions decay rapidly\cite{Yeh99}.
For pristine surfaces, due to the exponential decay with $z$ of the
electrostatic potential of the substrate and the absence of molecular
multipoles, this approach can be particularly efficient, allowing the
usage of a smaller vacuum padding region.

For an efficient implementation of PME, it is essential to optimally
split the charge density into a smooth (low\-/frequency) component
solved in the reciprocal space and a high\-/frequency residual handled
in the real space. This decomposition enables the use of relatively
coarse grids. Sophisticated smoothing and splitting schemes have been
developed for this purpose\cite{Arnold13}, but their details are
beyond the scope of this article. Nevertheless, even with these
optimizations (imposing PBCs on the system, using highly optimized PME
solvers and FFT libraries), electrostatics still accounts for a
relevant portion of the computational cost in classical molecular
dynamics simulations.

Here, GridFF offers major advantages in both speed and accuracy. Since
the electrostatic potential of the substrate is precomputed, the
computational cost of solving for the electrostatic potential (via
PME, or any other method) is removed entirely from the molecular
dynamics (MD) run. In addition, periodic boundary conditions are
imposed only on the substrate, and not artificially on the molecular
system too. This means that the molecule does not actually interact
with its periodic images.  Moreover, in fully periodic systems
(\emph{e.g.}, a pristine surface), all interactions can be folded into
the unit cell of the substrate, even if the molecular adsorbate is
much larger, and therefore storing all interaction data in a minimal
grid. During the MD run, the atomic coordinates of the adsorbate are
then mapped into the surface unit cell and used for computing energies
and forces. In addition, the surface potential becomes negligible just
a few \AA~above the substrate, as both Morse and electrostatic
components decay exponentially with the height. For example, the
potential of a pristine NaCl substrate can be stored in as little as
32 \AA$^3$ or 256 kB per component. For non\-/periodic systems
(\emph{e.g.}, point defects, step edges, AFM tips), the required
memory footprint is significantly higher, \emph{e.g.}, 100 cubic
nanometers and 800 MB per component in our largest supercell of
20$\times$20. Overall, aside from the assumption of surface rigidity, the
main limitation of GridFF is memory usage.

In the present work, we test the GridFF approach using a simplified
electrostatic solver, which, unlike PME, omits the real\-/space residual
entirely and relies solely on solving the Poisson equation $\nabla^2 V
= \rho$ in the reciprocal space. In this respect, our implementation
more closely resembles the electrostatic solvers used in density
functional theory codes (\emph{e.g.}, SIESTA\cite{Soler02} or
VASP\cite{Kresse96}), than those based on classical force fields like
LAMMPS. Besides simplicity, our motivation for this approach is the
possibility to generate GridFF directly from \emph{ab initio} charge
densities without requiring atomic charge assignment or
projection. Thanks to the relatively fine grid spacing (0.1 \AA),
this reciprocal\-/space\-/only solver achieves an accuracy of about
0.01 meV, which is sufficient for most purposes.  Nevertheless, in
future work, we plan to implement the full PME algorithm including the
real\-/space correction to further improve accuracy.

\section{Interpolation}\label{sec:interpolation} 

Among the many numerical algorithms developed for function
approximation and interpolation, we seek those that offer optimal
performance on current computing hardware while maintaining sufficient
accuracy. Additionally, we require methods that are completely general
(\emph{i.e.}, do not assume any particular functional form of the
potential) and that can be used not only for pairwise interaction
potentials, but also for potentials derived from electron
densities. These requirements effectively constrain us to low\-/degree
polynomial approximations defined on orthogonal uniform grids, which
are also compatible with FFT algorithms used in long\-/range
electrostatics.

A crucial consideration for high\-/performance implementation on
modern CPUs and GPUs is the memory access pattern and cache
locality. Contemporary hardware can perform hundreds of arithmetic
operations (\emph{e.g.}, additions or multiplications) in the time
that it takes to read a single number from global memory. To mitigate
this bottleneck, fast cache memory is used to pre\-/load data stored
at nearby addresses. Thus, interpolation algorithms can be
significantly accelerated when the data required to evaluate the
potential at a given atomic position are collocated in memory and
accessed in contiguous blocks.

In the original application of grid-projected potentials for AFM
imaging (ppafm)\cite{Hapala14, Oinonen2024}, the trilinear
interpolation was found sufficient, and even implemented in a
hardware\-/accelerated form with reduced numerical precision. However,
for simulations involving self\-/assembly or nanomanipulation of
molecules on surfaces (especially under low\-/temperature conditions),
a higher accuracy is required. Such systems can be sensitive to energy
differences on the order of fractions of a meV, which can
qualitatively affect trajectory bifurcations and final states.

Another fundamental problem of the trilinear interpolation approach
is the inconsistency between interpolated forces and energy. In that
implementation, the force components ($F_x$, $F_y$, $F_z$) and energy
$E$ were stored and interpolated independently. This not only require
storing four times more data in memory ($E$, $F_x$, $F_y$ and $F_z$,
instead of just $E$), but also implies that the resulting force is not
the exact gradient of the energy field. This is because the derivative
of a piecewise linear energy function is a piecewise constant
function, while the interpolated forces are instead piecewise
linear. This may cause difficulties to converge forces in molecular
dynamics below a certain threshold, and/or for energy conservation.

To resolve this, we implemented a tricubic interpolation of the
energy field, where forces are obtained analytically as the gradient
of the piecewise cubic polynomials. In one dimension, this produces
force profiles that are piecewise quadratic and therefore continuous
and smooth. Formerly, we tested several approaches, including Hermite
cubic polynomials, but ultimately selected cubic B\-/splines, which
provided the best trade\-/off between interpolation accuracy and
computational performance.  Tricubic B\-/splines are also used in some
ligand\-/docking applications, in cases where high quality forces are
required, although typically with coarser grid spacing\cite{Oberlin98,
  Trosset98}.

The one\-/dimensional cubic B\-/spline interpolation on an interval
$(x_i,x_{i+1})$ can be expressed as a linear combination
\begin{equation}
  f(x) = \sum_{j=-1}^{2} B \left( \frac{x-x_{i+j}}{\Delta x} \right) b_{i+j}
  \label{eq:Bspline1D}
\end{equation}
of B\-/spline basis functions $B(t)$ positioned at four points
($x_{i-1}$, $x_i$, $x_{i+1}$, $x_{i+2}$) separated by a grid spacing
$\Delta x$. The B\-/spline basis is a symmetric piecewise cubic
polynomial defined as
\begin{equation}
  B(t) = 
  \begin{cases}
    \frac{2}{3} - t^2 + \frac{1}{2}\,|t|^3, & |t| < 1, \\
    \frac{1}{6}\,(2-|t|)^3, & 1 \le |t| < 2, \\
    0, & |t| \ge 2.
  \end{cases}
  \label{eq:Bspline}
\end{equation}
The 3D cubic B\-/spline interpolation is implemented as a Cartesian
(tensor) product of 1D interpolations. This means that each evaluation
accesses data from a 4$\times$4$\times$4 block of the nearest grid points -- namely,
64 values in total. For each potential component (\emph{i.e.},
Coulomb, Pauli, and London), this corresponds to reading 192
floating\-/point numbers.
 
The grid is stored in memory such that the fastest\-/changing axis
(typically $z$) is laid out contiguously, so blocks of four values
along $z$ are collocated in memory.  Although the full interpolation
could be evaluated as a direct sum over 64 basis functions weighted by
precomputed coefficients, more efficient implementations -- like ours
-- decompose the operation into a sequence of 1D
interpolations. Specifically, we perform 16 1D interpolations along
$z$, followed by 4 along $y$, and finally 1 along $x$. Starting with
the fastest axis ($z$ in our case) ensures optimal cache locality and
allows the use of single\-/instruction\-/multiple\-/data
vectorization.

According to our benchmarks, this tricubic B\-/spline interpolation is
only about two times slower than a trilinear interpolation (using
grids for $E$, $F_x$, $F_y$ and $F_z$), while providing several orders
of magnitude higher accuracy. Moreover, in practical simulations, the
cost of evaluating molecule\-/substrate interactions (using either
interpolation scheme) is anyway about one hundred times faster (on
CPU) than the cost of evaluating molecule\-/molecule pairwise
interactions -- even for small systems ($\sim$50 atoms). Therefore,
the additional cost of the tricubic interpolation is negligible in
real workloads. On GPU, the GridFF interpolation consume a
significantly higher share of performance budget (see Section
\ref{fig:conf_sampling}) as the GPU performance is limited by global
memory access rather than arithmetic operations.

One complication of using B\-/splines is that the expansion
coefficients $b_i$, stored at each grid point, are not known \emph{a
priori}. They must be fitted such that the interpolated potential
reproduces the reference values at grid nodes. While this fitting is a
linear problem involving a very sparse matrix, the problem size is
large: for typical grid spacing ($\sim$0.1 \AA), a cubic nanometer
contains on the order of 10$^6$ grid points. This makes iterative
solvers more feasible than direct matrix inversion approaches.

Currently, we use a simple gradient descent algorithm to minimize the
root mean square error across all grid points. Although this method is
not particularly efficient and may require thousands of iterations to
reach convergence at the desired accuracy, we do not consider this as
a bottleneck. In a typical workflow, the grid is precomputed once for
a given substrate, and then reused in many simulations.  Therefore, we
have not prioritized further optimizations of the coefficient fitting
procedure.

\section{Accuracy and performance tests on CPU} \label{sec:performance}

In order to benchmark the accuracy of the proposed approach, we
performed several systematic tests calculations using GridFF as
implemented in FireCore\cite{FireCore} and all\-/atom simulations
using the LAMMPS package\cite{Thompson22}. The studied system is the
desorption and manipulation of a 3,4,9,10\-/perylenetetracarboxylic
dianhydride (PTCDA) molecule on top of a NaCl slab. This system was
chosen as it has been extensively studied experimentally by means of
SPM techniques\cite{Karacuban11, Swart12, Decampos24}. Also robotic
manipulation of the PTCDA molecule (although on metallic substrate)
originally motivated our work \cite{Leinen20}.

First, we performed rigid (\emph{i.e.}, with fixed relative positions
of the atoms in the molecule) vertical and lateral scans of PTCDA on a
8$\times$8$\times$3 NaCl(001) surface slab to compare individual potential
components between GridFF and LAMMPS.  Next, we conducted relaxed
scans with the same molecular and substrate configurations, allowing
the system to undergo molecular relaxation, therefore more closely
mimicking a manipulation experiment with AFM. After validating the
relaxed potential behavior, we extended our tests to include defective
substrates. We introduced a neutral, nearly isolated defect by
removing 2 atoms from the surface of a 20$\times$20$\times$3 NaCl supercell,
equivalent to a defect density of approximately 0.08\%. Using the same
PTCDA molecule, we performed lateral scans across this defected
surface to test the capability of the method to handle surface
irregularities. Finally, we compared the performance of GridFF and
all\-/atom calculations for increasing substrate sizes ranging from
8$\times$8$\times$3 to 20$\times$20$\times$3 supercells.

The molecular structure of PTCDA was retrieved from a previous work of
one of us on high resolution AFM imaging\cite{Oinonen2024}, and
supplemented with atomic charges determined using the restrained
electrostatic potential fitting scheme\cite{Bayly93}. The NaCl
substrate was modeled by three \emph{fcc} atomic layers with lattice
spacing of 4.0 \AA, and charges for the Na/Cl ions set to $\pm$0.9e
according to the Bader analysis\cite{Bader90}. The intramolecular
interactions are modeled using the universal force field
(UFF)\cite{Rappe92} both in FireCore and LAMMPS. UFF values were also
used to obtain the Morse parameters modeling the molecule\-/substrate
interaction (the value of the $\alpha$ parameter was set to 1.5
\AA$^{-1}$ for all pairs, and a cutoff of 17 \AA~was applied).

Within the GridFF framework, we generated potential energy grids for
the NaCl(001) substrate using a fine grid spacing of 0.1 \AA~and
setting the maximum iterations to 3000 for the generation of the
B\-/spline parameters, achieving fitting errors on the order of
10$^{-5}$ for both Morse and Coulomb potentials. For the treatment of
the slowly decaying electrostatic interactions, in FireCore we used a
reciprocal\-/space Poisson solver as reported above, whilst in LAMMPS
they were evaluated with the closely related particle\-/particle
particle\-/mesh (PPPM) method\cite{Hockney88} with a real space cutoff
of 15 \AA. The relative accuracy of reciprocal part of PPPM kernel was
set to 10$^{-8}$ in rigid scans to obtain high-accuracy reference
energy profiles.  Such a stringent tolerance was necessary since a
more standard setting (10$^{-6}$) produced numerical artifacts in
LAMMPS calculations larger than the error coming from the
reciprocal-space-only Poisson solver in GridFF.  Nevertheless, a more
relaxed threshold of 10$^{-6}$ was chosen for relaxed scans with
LAMMPS, as these simulations were also used for benchmarking the
performance in typical use\-/cases.  Therefore we should note that in
case of relaxed scans FireCore simulations are actually more accurate
then LAMMPS results, while being significantly faster.  To ensure a
fair comparison, both FireCore and LAMMPS used the same FIRE
algorithm\cite{Bitzek06} for the structural optimization with a
convergence threshold for forces of 10$^{-3}$ eV/\AA.

\subsection{Rigid scans of PTCDA on NaCl} \label{sec:rigid}

\begin{figure}[htbp]
  \centering
  \includegraphics[width=\linewidth]{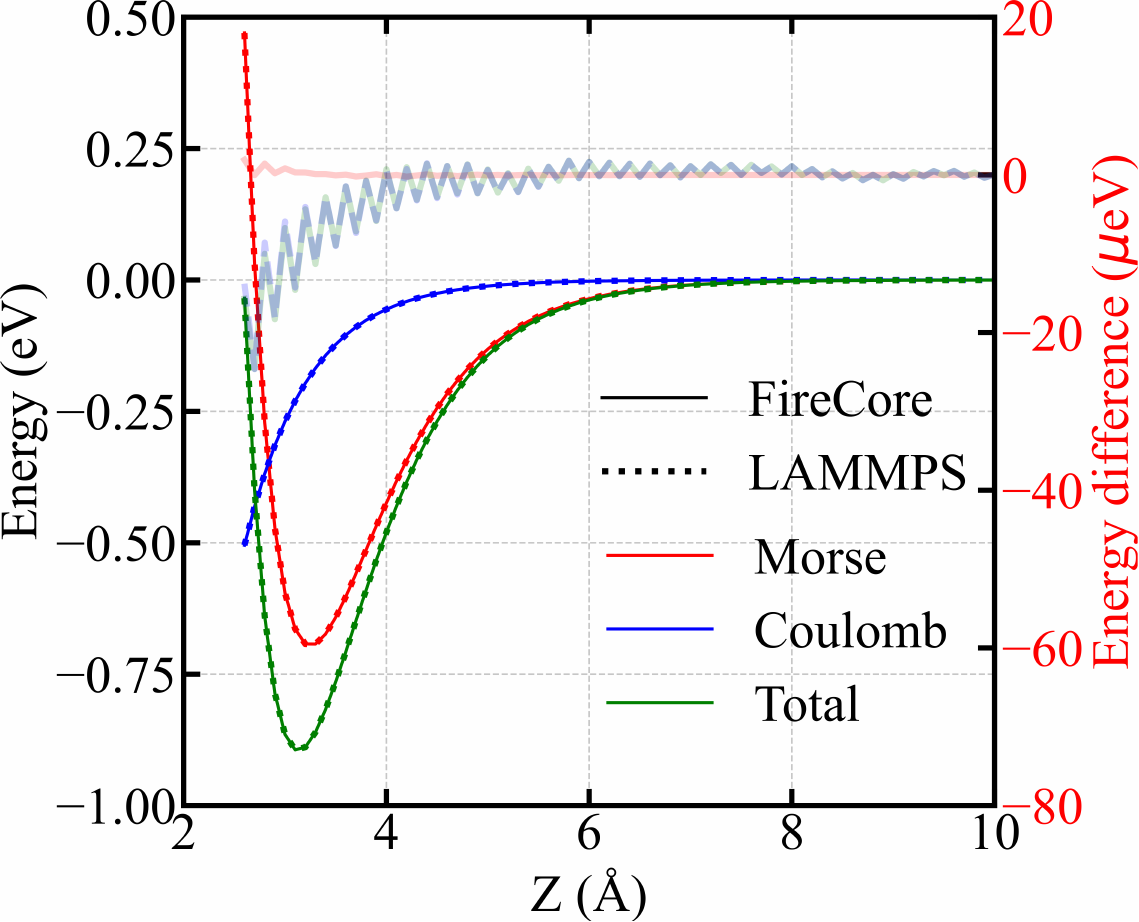}
  \caption{Energy profiles for a PTCDA molecule interacting with a
    NaCl surface as a function of the separation distance along the
    $z$-direction. The total potential is decomposed into Morse (red)
    and Coulomb (blue) components, with the total energy (green) shown
    for the sake of completeness. Calculations using the GridFF
    approach as implemented in the FireCore code are reported with
    thin solid lines, whilst all\-/atom simulations from LAMMPS are in
    thick dashed lines. The energy differences between the two methods
    are plotted with semi\-/transparent lines and with the same color
    scheme.}
  \label{fig:rigid_z}
\end{figure}

The profiles in Figure \ref{fig:rigid_z} show the contributions to the
energy variations when a PTCDA molecule is rigidly moved away from the
NaCl surface. The decomposition of the total interaction potential
into individual components ensures that the underlying interaction
physics is accurately reproduced by the GridFF approach. The Morse
potential component (red) describes the short\-/range interactions
(encompassing both Pauli repulsion and van der Waals attraction),
whilst the Coulomb potential component (blue) captures the
long\-/range electrostatics between the PTCDA molecular charge
distribution and the ionic NaCl substrate. The total potential energy
curve (green) represents the superposition of all interaction
components, yielding an equilibrium adsorption configuration at Z =
3.1 \AA~with a total binding energy of -0.89 eV. All energy profiles
overlap perfectly, and therefore, to quantitatively estimate the
accuracy of GridFF approach, we also report the energy differences
between the two methods on the right $y$-axis. In all cases, the
agreement is remarkable (considering the fitting error in the
interpolation procedure, and the PPPM accuracy set in the LAMMPS
calculations), with the Morse potential component exhibiting
differences on the order of 10$^{-6}$ eV, whilst the Coulomb component
shows discrepancies of approximately 10$^{-5}$ eV. Overall, the
differences are dominated by the electrostatics contribution, due to
the plane\-/wave cut\-/off of the reciprocal Poisson solver and the
aforementioned omission of real\-/space residual from PME employed in
the GridFF calculations. In both cases, the differences tend to
increase at small separation distances (where the slope of the
profiles is maximal in absolute values).

\begin{figure*}[htbp]
  \centering
  \includegraphics[width=\linewidth]{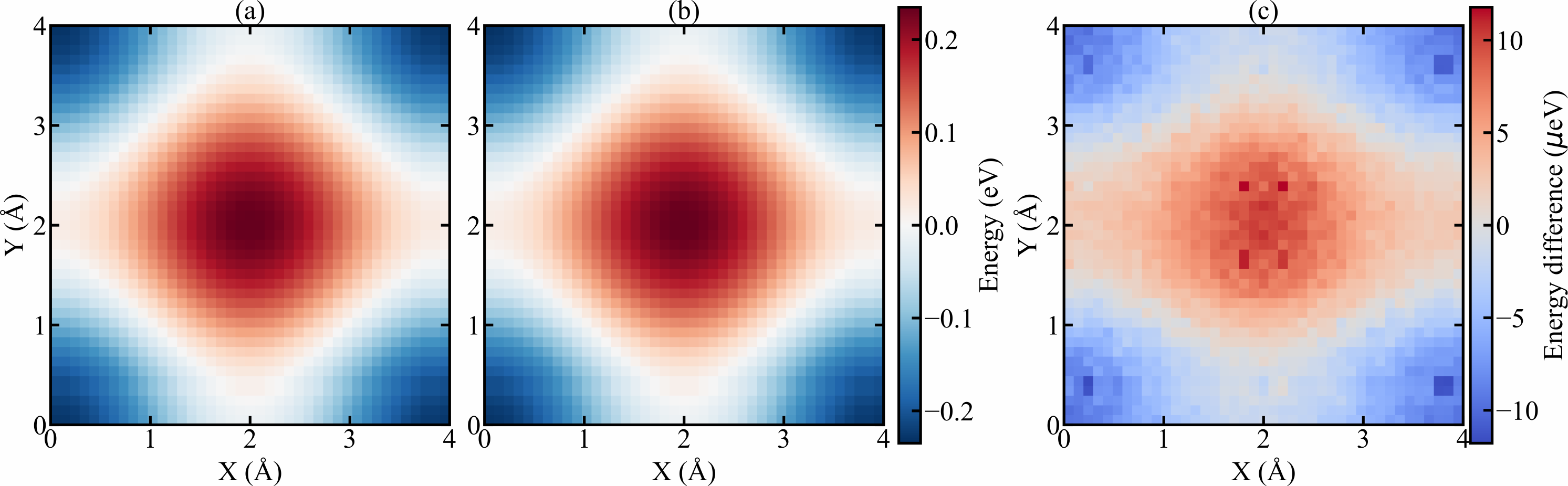}
  \caption{Rigid PES of a PTCDA molecule interacting with a NaCl
    surface at a separation height of 3.3 \AA, obtained with the
    GridFF approach as implemented in the FireCore (a), and all\-/atom
    calculations from LAMMPS (b). Panel (c) reports the energy
    difference between FireCore and LAMMPS.}
  \label{fig:rigid_xy}
\end{figure*}

Next, we performed lateral two\-/dimensional rigid scans on the
$xy$-plane by rigidly displacing the PTCDA molecule with a step length
of 0.1 \AA~and at a constant height of 3.3 \AA~from the
surface. Figure \ref{fig:rigid_xy} shows the total interaction energy
as a function of the in\-/plane position of the molecule across one
unit cell of the NaCl(001) substrate. Such a 2D scan provides
information about the preferred adsorption sites and barriers for
lateral diffusion or manipulation of the molecule on the surface.  The
left and center panels show the PES calculated by FireCore and the
reference LAMMPS force field, respectively. Both methods produce a
qualitatively identical energy landscape, with an energy barrier of
0.46 eV located at the center of the unit cell. The right panel
displays the absolute energy difference between the FireCore and
LAMMPS calculations. As in the previous case, the absolute error is on
the order of 10$^{-5}$ eV. In summary, these results demonstrate that
FireCore accurately reproduces the whole interaction energy landscape.

\subsection{Relaxed detachment of PTCDA from NaCl} \label{sec:relax}

\begin{figure*}[htbp]
  \centering
  \includegraphics[width=\textwidth]{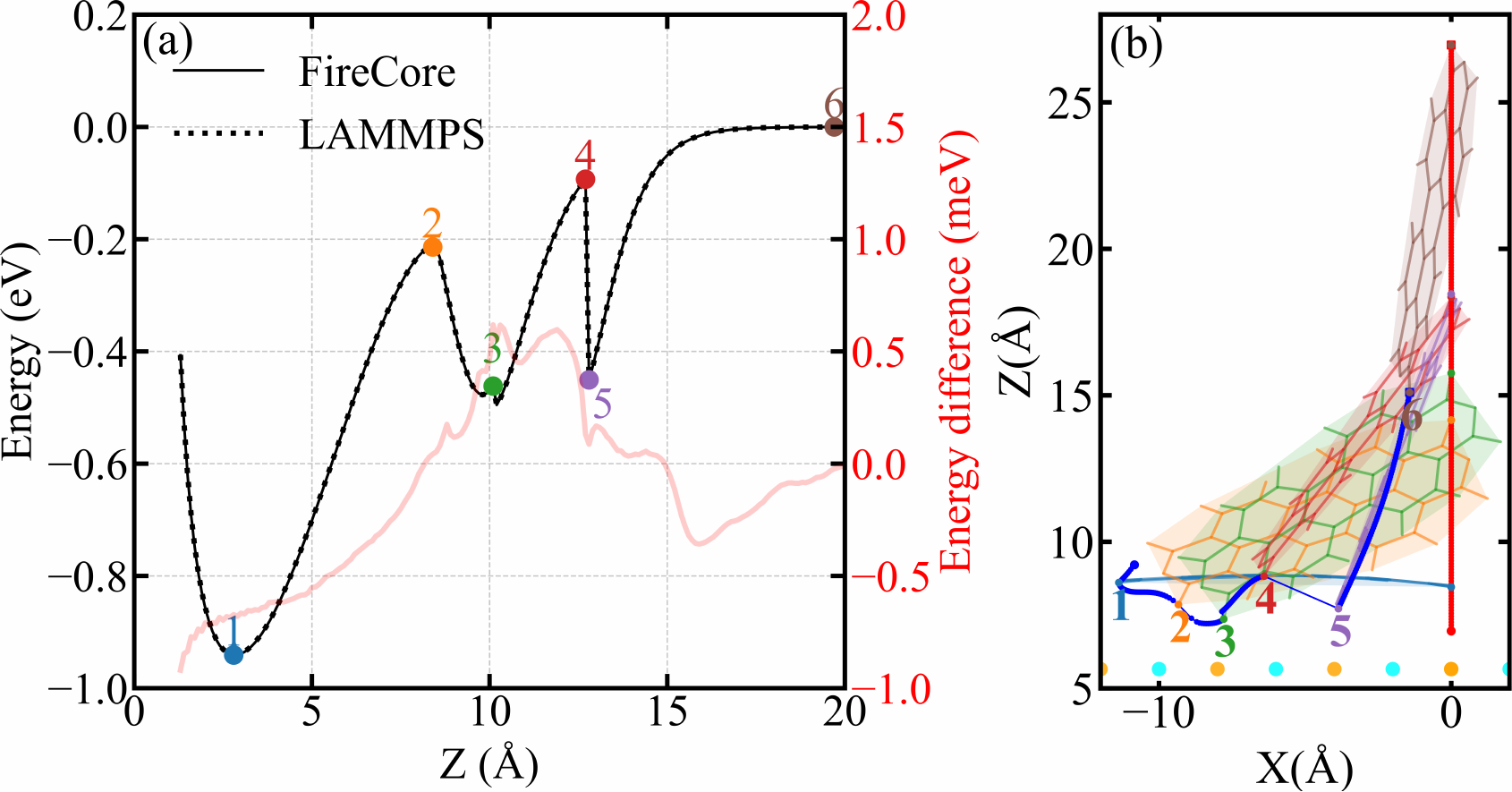}
  \caption{(a) Energy profiles for both GridFF (FireCore) and
    all\-/atom (LAMMPS) calculation, and (b) selected configurations
    of a PTCDA molecule lifted up from a NaCl surface. In panel (b),
    the red and blue dots represent the atom that is displaced
    vertically during the scan, and the carbonyl oxygen at the
    opposite position of the PTCDA molecule, respectively. The six
    molecular configurations shown in (b) correspond to the marked
    dots in (a). Na and Cl ions are depicted in orange and cyan,
    respectively (only the topmost layer is shown).}
  \label{fig:relax_scan}
\end{figure*}

In order to simulate more realistically the manipulation of PTCDA with
an SPM tip on a surface at low temperature, we performed calculations
where we restrained the position of one atom (a carboxylic oxygen at
the corner) of the molecule and let the position of the other atoms
relax. In the first set of simulations, at each step we displaced the
$z$-component of the position of the selected atom by 0.1 \AA,
starting from 1.3 to 20 \AA~above the NaCl(001) surface. Such a
computational setup effectively mimics the experimental protocols
where a functionalized tip is used to grasp and manipulate individual
molecules on surfaces by forming a mechanical contact with a specific
atomic site\cite{Scheidt23, Leinen20}. The resulting relaxed potential
energy curve reported in Figure \ref{fig:relax_scan} exhibits
significantly different characteristics compared to the rigid
scan. The global minimum occurs at approximately Z = 2.8 \AA~with a
binding energy of -0.94 eV, representing the equilibrium adsorption
configuration. The larger binding energy (compared to the rigid scan)
demonstrates the importance of molecular flexibility in achieving
optimal molecule\-/surface interactions through conformational
adaptation. The unbinding energy profile is also qualitatively
different, presenting multiple local minima and discontinuities,
particularly in the intermediate separation range (6-15 \AA). These
features arise from the interplay between attractive
molecule\-/surface interactions and internal molecular strain. The
molecular relaxation allows for rotation, tilting, and conformational
changes (that can create metastable adsorption states not accessible
in rigid scan calculations), and eventually leading to the sudden
detachment of the molecule from the substrate. As already pointed out,
in the proximity of the sudden changes of molecular
orientation/conformation, the relaxation dynamics is more sensitive to
small force differences, leading to bifurcation in the taken
path. Nevertheless, also in this case the quantitative agreement
between FireCore and LAMMPS is good, with energy differences remaining
below 0.9 meV across the entire scan range. These results validate the
ability of the GridFF approach to accurately capture the energetics
and the structural response of a molecule under mechanical
manipulation, a critical requirement for predicting AFM\-/based
molecular device fabrication and single\-/molecule manipulation
protocols.

\subsection{Dragging the PTCDA molecule over a defect}

\begin{figure*}[htbp]
  \centering
  \includegraphics[width=\textwidth]{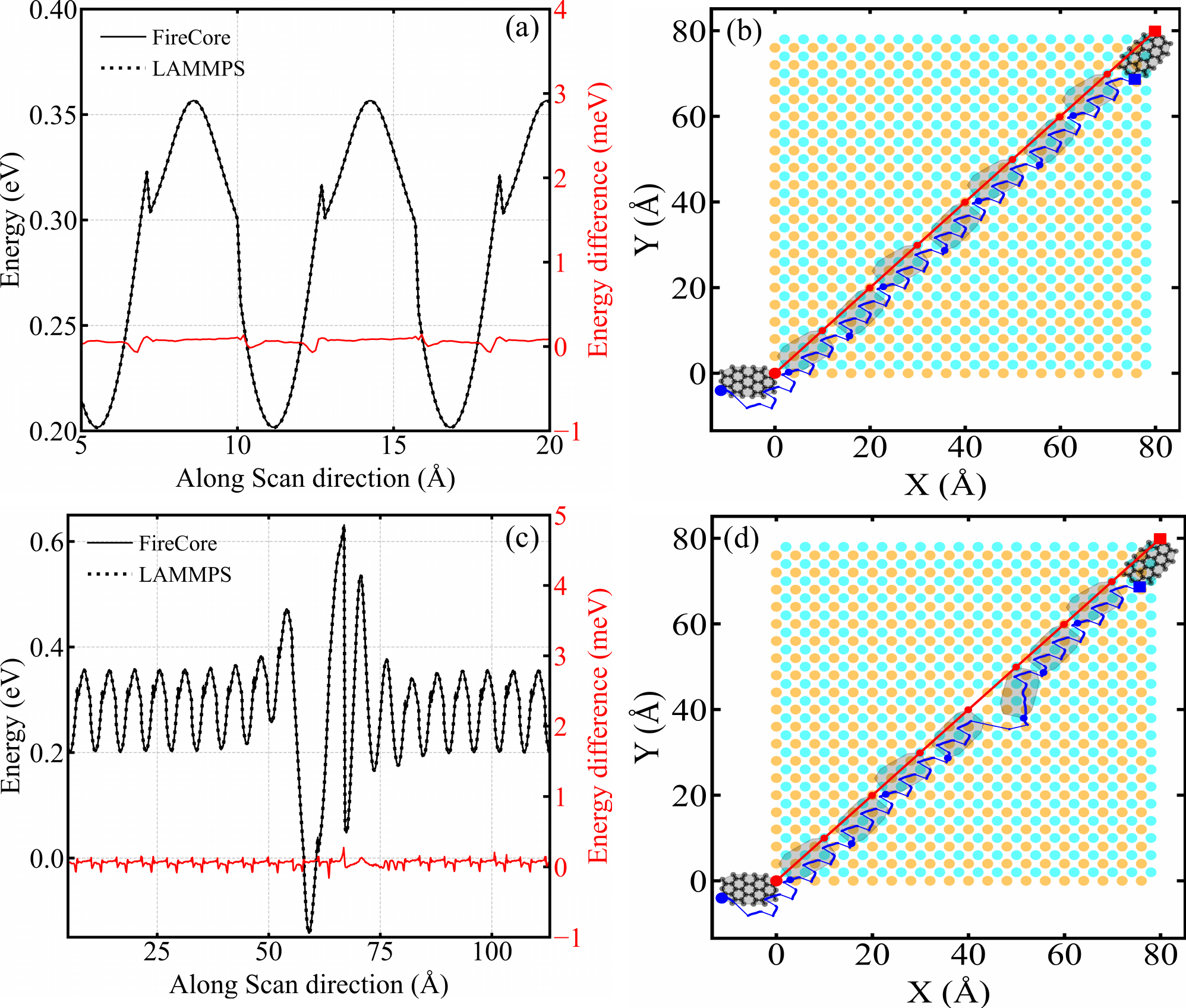}
  \caption{Comparison between GridFF (Firecore) and all\-/atom
    (LAMMPS) calculations for a PTCDA molecule dragged over a NaCl
    surface with and without the presence of a defect. (a) Energy
    profile and (b) the corresponding set of relaxed configurations on
    a pristine 20x20x3 NaCl substrate. (c) Energy profile and (d) the
    corresponding set of relaxed configurations on the same substrate
    in the presence of an isolated neutral defect at the center of the
    cell. In the energy plots (a, c), the potentials calculated by
    FireCore (dashed orange) and LAMMPS (solid blue) are shown on the
    left axis, with their absolute difference (red) on the right
    axis. In the trajectory plots (b, d), the path of the restrained
    atom (red) and of the atom in the opposite corner of the molecule
    (blue) are shown with dots, illustrating the path taken by the
    molecule over the Na (orange) and Cl (cyan) ions of the top
    substrate layer.}
  \label{fig:all_trajectories}
\end{figure*}

The goal of this set of simulations is to mimic the dragging of a
PTCDA molecule by a SPM molecule over a NaCl surface along the
diagonal of the substrate cell. As done before, this was achieved by
fixing the position of one atom of the PTCDA molecule (marked by the
red dot in Figure \ref{fig:all_trajectories}), and systematically
displacing it along the scan direction while allowing the entire
molecular system to relax at each step. We have considered a 20x20x3
NaCl substrate with 2400 atoms for the pristine surface, and a system
with a neutral defect (\emph{i.e.}, by removing one Na and one Cl
atoms next to each other from the topmost atomic layer) placed at the
center of the cell.

The top panels of Figure \ref{fig:all_trajectories} illustrate the
sliding behavior on a defect\-/free NaCl surface. The energy profile
in Figure \ref{fig:all_trajectories}a shows a quantitative agreement
between FireCore and LAMMPS, with the difference remaining
consistently below 0.9 meV. The energy minima correspond to the PTCDA
molecule settling into energetically favorable adsorption sites that
align with the underlying Na and Cl ion lattice. The energy maxima
represent the potential barriers that the molecule must overcome to
move between these stable sites. The non\-/smooth nature of the energy
profile arises from the interplay between the static interaction field
generated by the substrate and the orientational and conformational
degrees of freedom of the PTCDA molecule. As it is pulled across the
surface, the molecule continuously adjusts its orientation and
internal geometry to minimize the total energy, leading to the
characteristic 'stick\-/slip' motion (reminiscent of different systems
studied by AFM experiments and MD simulations\cite{Ouyang18,
  Vilhena22}) visually depicted in the trajectory plot (Figure
\ref{fig:all_trajectories}b). The path of the unconstrained opposite
corner atom (blue dot) clearly deviates from the straight\-/line path
of the fixed atom, hopping between adjacent lattice sites.

The bottom panels of Figure \ref{fig:all_trajectories} shows the
capability of FireCore to handle localized and chemically complex
features, such as a neutral point defect in the substrate. It can be
observed from the graph that, far from the defect, the energy profile
in Figure \ref{fig:all_trajectories}c retains the periodic corrugation
of the pristine surface. However, as the molecule approaches the
defect location (at the mid\-/point of the scan length), the energy
profile changes drastically. A sharp, deep potential well emerges,
indicating a strong pinning of the molecule to the defect site. This
interaction is significantly stronger than the regular surface
corrugation, with an energy stabilization of over 0.77 eV. Figure
\ref{fig:all_trajectories}d provides an intuitive real\-/space
visualization of this event: the path taken by the molecule shows a
lateral deviation as it is influenced by the presence of the
defect. The PTCDA molecule reorients itself to maximize its favorable
interaction with the defect before being pulled away, which requires
overcoming a substantial energy barrier. Also in this case, the energy
profile obtained with GridFF perfectly matches the one from reference
calculations, with maximum deviations in the order of 0.9
meV. Overall, this demonstrates the capability of the proposed
approach to accurately model the molecule\-/surface interaction even
in the presence of more complicated features such as a point defect,
making it a powerful tool for predictive materials simulation.

\subsection{Performance comparison}

The lateral relaxed scan on the pristine surface described in the
previous section was also used to benchmark the performance of the CPU
implementation of GridFF in FireCore with respect to LAMMPS.  The
dynamical relaxation with the FIRE algorithm\cite{Bitzek06} is a good
choice for such comparison, as the overheads (\emph{e.g.}, setup and
initialization of the system) are amortized over the thousands of
steps required for the structure relaxation.  To ensure a reliable
comparison between the two approaches, and avoid any dependence on the
details of the relaxation algorithm, the number of relaxation steps
was set to the same value (5000), and with a tiny threshold on the
force convergence criterion to ensure that the maximum number of
relaxation steps is always reached.  In this way we can directly
compare the calculation walltimes needed to complete the dragging
path, as in both cases the same number of relaxation steps (and
therefore of force and energy evaluations) are performed. It must be
stressed that, in order to have a fair comparison, the PPPM accuracy
tolerance in LAMMPS was increased to 10$^{-6}$ (with respect to
10$^{-8}$ used for rigid-scan calculations).  Moreover, again for the
sake of fairness, we performed calculations also directly excluding
the computation of pairwise interactions within the substrate, by
removing the appropriate atoms from the neighbor lists.

\begin{figure}[htbp]
  \centering
  \includegraphics[width=\linewidth]{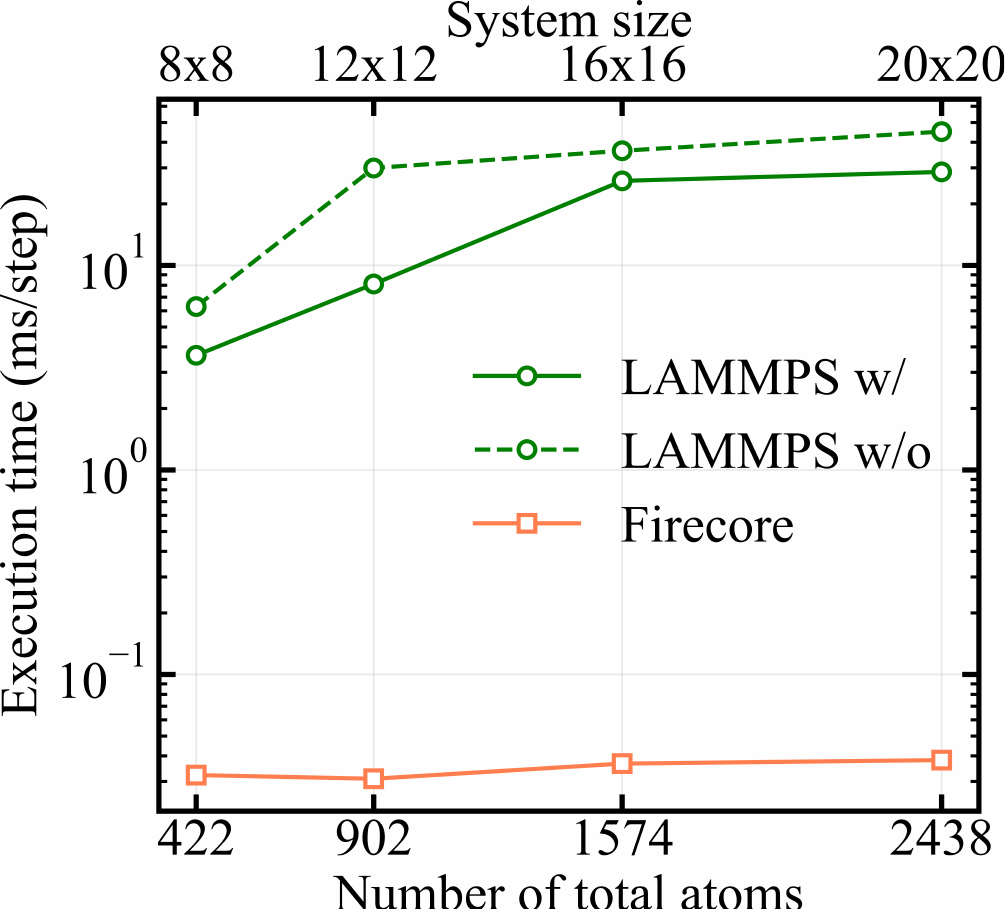}
  \caption{Execution walltimes for FireCore and LAMMPS as a function
    of the system size. The green solid and dashed lines represent
    LAMMPS calculations with and without excluding the computation of
    pairwise interactions within the substrate atoms, respectively.}
  \label{fig:CPU_perfor}
\end{figure}

Figure \ref{fig:CPU_perfor} shows the comparison of the computational
performance between LAMMPS and Firecore, together with the scaling
with respect to the system size (ranging from 8$\times$8 to 20$\times$20 NaCl unit
cells, corresponding to 384 to 2400 substrate atoms). All calculations
were performed on the same computer equipped with an AMD EPYC 7513 CPU
(2.6 GHz) using a single core. Firecore achieves good speedup factors
ranging from 113$\times$ for the smallest system to 751$\times$ for the largest,
reducing execution times from hours to minutes, as also shown in Table
\ref{tab:cpu_performance_comparison}. Notably, the largest system,
that requires over 45 hours in LAMMPS, completes in less than 4
minutes with Firecore. Moreover, as one can notice from the different
slopes in Figure \ref{fig:CPU_perfor}, GridFF shows also a superior
scaling behavior with respect to the system size, if compared to
all\-/atom calculations. In fact, by moving from the smallest to the
largest systems, the total number of atoms increases by about 6
times. Correspondingly, the execution time (normalized by the total
number of path points) increases by almost 8 times in LAMMPS, while in
FireCore the increment is only about 20 \%.  Such dependence of speedup
factors on the system size indicate that the algorithmic advantages of
GridFF may become even more pronounced for larger systems (such as
pulling of graphene ribbons \cite{Kawai16} or DNA), enabling practical
high\-/throughput screening and statistical sampling for several
applications.

\begin{table*}[htbp]
  \centering
  \caption{Execution times and computational speedups for FireCore
    and LAMMPS for different system sizes.}
  \label{tab:cpu_performance_comparison}
  \begin{tabular}{c|c|c|cccc}
    \hline
           &         &          & \multicolumn{4}{c}{LAMMPS} \\
    System & Total   & Firecore & \multicolumn{2}{c|}{with exclusion} & \multicolumn{2}{c}{without exclusion} \\
    size   & steps   & Time (s) & \multicolumn{1}{c|}{Time (s)} & \multicolumn{1}{c|}{Speedup} & \multicolumn{1}{c|}{Time (s)} & Speedup \\
    \hline
    8x8    & 2265000 & 73       & \multicolumn{1}{c|}{8240}     & \multicolumn{1}{c|}{113}        & \multicolumn{1}{c|}{14256}    & 195  \\
    12x12  & 3395000 & 105      & \multicolumn{1}{c|}{27549}    & \multicolumn{1}{c|}{262}        & \multicolumn{1}{c|}{101655}   & 968  \\
    16x16  & 4525000 & 166      & \multicolumn{1}{c|}{117259}   & \multicolumn{1}{c|}{706}        & \multicolumn{1}{c|}{164526}   & 991  \\
    20x20  & 5660000 & 216      & \multicolumn{1}{c|}{162257}   & \multicolumn{1}{c|}{751}        & \multicolumn{1}{c|}{255390}   & 1182 \\
    \hline
  \end{tabular}
\end{table*}


\section{Configuration sampling on GPU}\label{sec:config_sampling}

While the simulations of dragging of the PTCDA molecule using CPU
allows a side\-/by\-/side comparison of the accuracy and performance
of GridFF/FireCore with LAMMPS, the main strength of our approach lies
in its ability to accelerate the exploration of large configuration
spaces of flexible molecules on surfaces.  This includes tasks such as
finding the most stable binding configuration (\emph{i.e.}, the global
energy minimum in the configuration space, which is a hard problem),
or computing the binding potential of mean force, which requires
sampling all energetically relevant configurations.

To illustrate the performance of FireCore in such applications, we
conducted a case study of the adsorption of a xylitol molecule on a
sodium chloride surface with a single chlorine vacancy.  The presence
of five hydroxyl groups in the xylitol molecule creates numerous
possibilities for hydrogen bonding with the ionic surface. Combined
with molecular flexibility due to free rotations around sigma bonds,
this results in a complex energy landscape characterized by many local
minima.

To comprehensively explore this vast conformational space, we employed
a minima hopping technique adapted for surface adsorption. This method
systematically samples different configurations by perturbing the
molecular structure (\emph{i.e.}, by performing 1000 steps of Langevin
molecular dynamics at a relatively high temperature of 300 K, followed
by dynamical relaxation to the nearest local minimum.  Each energy
minimization was carried out until forces converged below 0.1 meV/\AA,
ensuring an accurate representation of the stable configuration.

Efficient implementation of molecular dynamics for small molecules
like this on GPU faces several challenges.  Modern GPUs are equipped
with thousands of cores, which is significantly more than the number
of atoms in such systems (typically 50-100 atoms; rigid substrate
atoms represented by GridFF are excluded).  Although parallelization
over individual pairwise interactions (rather than atoms) is possible,
synchronized output of forces (\emph{i.e.}, reduction) from different
threads to global memory would require thread synchronization,
reducing performance.  We address this challenge by simulating
multiple replicas of the same system in parallel.  Testing showed that
the optimal performance is achieved with $\sim$5000 replicas, in the
case of the xylitol molecule.  Moreover, the time required to evaluate
forces (\emph{i.e.}, summing over all bonding and non\-/covalent
interactions) for such small systems is often shorter than the time
required to transfer atomic coordinates (and forces) to and from the
GPU. Therefore, the entire molecular dynamics loop -- including force
evaluation and integration of the equations of motion -- must be
executed entirely on the GPU, eliminating the need for costly
synchronization with the CPU.

While Langevin dynamics is performed fully on the GPU and downloaded
only every few hundred steps for visualization, the dynamical
relaxation requires global properties which involve reductions over
all atoms -- \emph{i.e.}, thread synchronization.  In particular, for
dynamical relaxations, one needs to calculate the norm of the whole
force vector $\vec F$ in order to check the force convergence
criterion, and to set to zero all velocities ($\vec v=0$) if the
system inertially moves up the hill ($\langle \vec{F} \mid \vec{v}
\rangle<0$).  To handle this, we download the system state every 100
steps and perform these operations on the CPU (although in principle,
such reductions could also be performed on the GPU but at the cost of
more complicated kernels).  At any rate, not performing the check at
every step does not seem to significantly hamper the overall
performance of the algorithm. We attribute this to the smoothness of
the trajectory near the minimum (which often is a long narrow
``valley'') and to the fact that a relaxation typically takes several
thousand steps anyway.

To identify unique structures, each optimized geometry is compared
with all previously minimized structures using a root-mean-square
deviation criterion (with a threshold of 0.1 \AA), accounting
for both conformational differences and variations in adsorption site
and orientation relative to the surface.  For this purpose, geometries
are downloaded from the GPU and the comparison is performed on the
CPU, typically only once per several thousand steps. Although the
computational cost of comparison with hundreds or thousands of
previously found local minima is substantial (using only CPU), it is
amortized in the dominating cost of thermalization and relaxation, and
therefore we did not attempt to implement this sub-task on GPU yet.

\begin{figure*}[htbp]
  \centering
  \includegraphics[width=\textwidth]{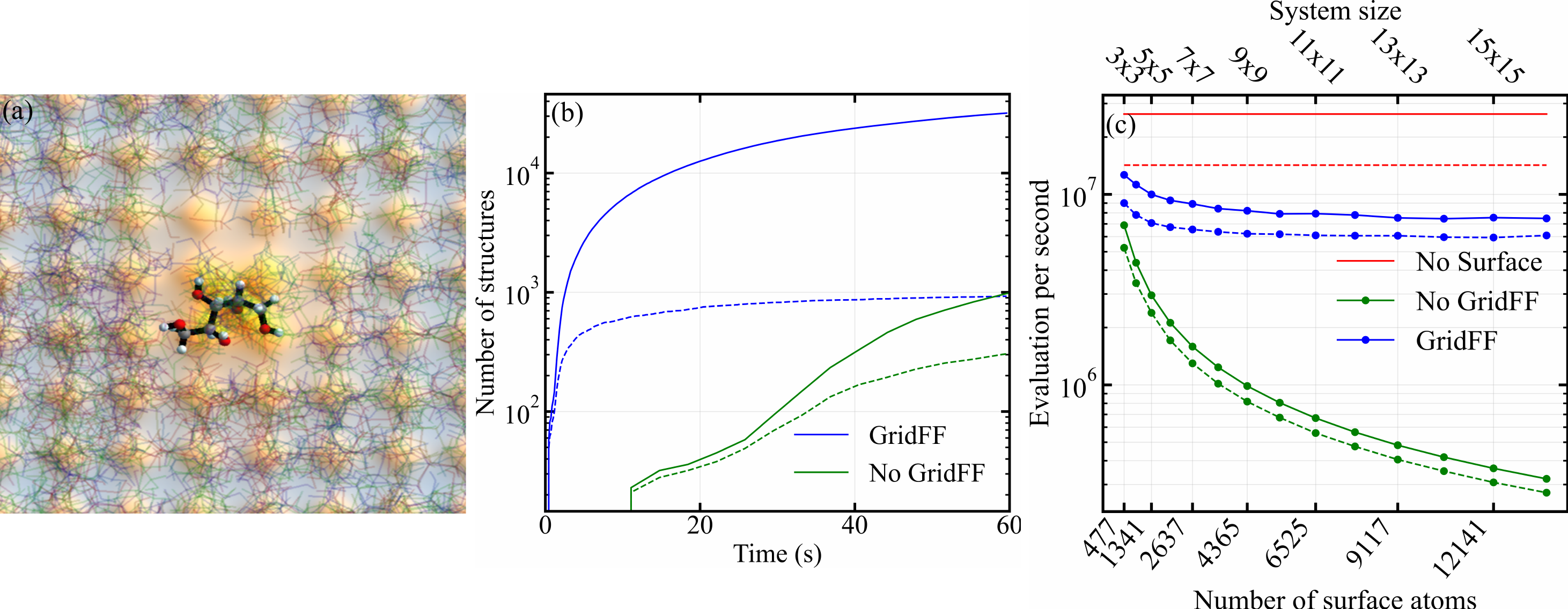}
  \caption{Analysis of xylitol adsorption on NaCl using GridFF. (a)
    Visualization of a 2000-replica simulation of representative
    xylitol molecule configurations adsorbed on the NaCl(001) surface
    with a Cl\-/hole.  (b) Number of total (solid) and unique (dashed)
    structures found in a short run using FireCore. Blue and green
    lines represent the outcomes of simulations done on
    16$\times$16$\times$3 surface with a Cl\-/hole using the GridFF
    and standard all\-/atom approaches, respectively. (c) Number of MD
    steps per second using GridFF (blue) and all\-/atom (green)
    calculations, and for an isolated xylitol molecule (red) for
    reference, as a function of the system size.  Solid and dashed
    lines corresponds to calculations with different settings.}
  \label{fig:conf_sampling}
\end{figure*}

The simulation can be run either as a batch calculation via the
FireCore Python API -- suitable for supercomputer production runs --
or through a graphical user interface (GUI), useful for debugging and
educational purposes on desktop computers.
Figure~\ref{fig:conf_sampling}a shows a screenshot from such an
interactive simulation, highlighting a selected configuration near the
vacancy, while the other replicas are shown as transparent skeletons
in the background.

In Figure~\ref{fig:conf_sampling}b,c we present performance metrics
for this simulation using both the GridFF approach and a naive
all-atom FireCore GPU MD-loop implementation. In the all-atom
simulation, molecule\-/substrate interactions are computed as a direct
sum of pairwise interactions (\emph{i.e.}, without Ewald summation as
used in LAMMPS).  These are summed over the nearest periodic images of
the surface supercell, while the interactions between surface atoms
are omitted.  Figure~\ref{fig:conf_sampling}c reports the total number
of substrate atoms interacting with the molecule (\emph{i.e.}, the
number of atoms in the supercell plus its 9 periodic images used for
the direct\-/sum evaluation).  Notably, GridFF-based calculations are
not only significantly faster but also more accurate, owing to the use
of Ewald summation in constructing the GridFF.

Figure~\ref{fig:conf_sampling}b presents a quantitative analysis of
the sampling efficiency, showing the number of converged and unique
structures found as a function of the execution time. The number of
converged structures increases proportionally with the number of
attempts, indicating the consistent performance of the energy
minimization protocol. In contrast, the number of unique structures
rises rapidly at early times and plateaus around 800 structures,
despite continued sampling.  This saturation suggests that our
approach exhaustively explored the relevant conformational space of
the xylitol\-/NaCl system under the chosen computational model.

Looking closely at Figure~\ref{fig:conf_sampling}b, we observe that,
in the all-atom simulation, the first batch of structures is found
only after $\sim$10 seconds, reflecting the time required for the
initial thermalization.  This synchronization delay also explains why
the green curve deviates from a steady\-/state sampling behavior
around 20 seconds after the simulation start.  A similar behavior is
seen in the GridFF simulation, but it is compressed into the first 1-2
seconds due to its $\sim$20$\times$ higher performance.

Figure~\ref{fig:conf_sampling}c provides a systematic analysis of how
performance scales with system size.  GridFF shows clear performance
superiority for larger supercells, achieving a throughput of
approximately 7.5 million MD steps per second for the largest systems,
compared to only 0.3 million steps per second for the all-atom
approach. Notably, GridFF performance remains stable regardless of the
system size, while the all-atom approach degrades steadily with system
size, making the performance advantage of GridFF even more pronounced
for larger systems.

For small supercells ($<$200 atoms in the supercell, $<$1800 atoms in
9 images), GridFF shows a slight increase in performance, likely due
to an improved cache efficiency as neighboring voxel memory addresses
are more contiguous. This effect could potentially be exploited
further by using coarser grids.

\section{FireCore from the user perspective}\label{sec:Firecore}

FireCore is written in C++ and OpenCL languages, with python binding
providing a convenient scripting interface. The generation of the
surface potential can be done internally on CPU, or using
a standalone pyOpenCL accelerated python script, which takes the
structure of the substrate (provided by the user in the \texttt{xyz} or \texttt{mol}
format), and the desired voxel size for the grid, and it generates and
stores the grid potential. The syntax for generating the grid
potentials presented in this paper is:
\begin{verbatim}
  python3 generate_grid.py
\end{verbatim}
The main simulation engine of FireCore can perform energy optimization
or molecular dynamics calculations of molecules on surfaces. To start
a run, is sufficient to provide the geometries of the molecule and the
substrate. Supported formats are \texttt{xyz} and \texttt{mol}. In the
case that a plain \texttt{xyz} file is provided for the molecule,
FireCore automatically detect the bonding topology.  FireCore
accordingly assigns the atomic types and all parameters needed for
bonded and non\-/bonded interactions.  Currently are available only
UFF and sp3FF (our home brew potential which was not yet published).
The user can also directly customize the parameter files for bonded
and non\-/bonded interactions. Next, the user can decide whether to
run grid\-/based or all\-/atom simulations. A folder (which also
contains a \texttt{README} file with commands and explanations) with
all files needed to run the examples reported in this paper is
available in the FireCore repository\cite{GridFF}.  To reproduce the
calculations presented in Section \ref{sec:performance}, one can run
the following command:
\begin{verbatim}
  python3 generate_scans.py 
\end{verbatim}

Along with the command line interface, FireCore also provides a GUI
for running interactive simulations, real\-/time visualization of
structural relaxations and molecular dynamics runs, including the
multiple\-/replica feature described in Section
\ref{sec:config_sampling}. The GUI allows the user to visualize atomic
types, charges, and electrostatic potentials. It also enables the
interactive manipulation of molecules by picking and dragging atoms
with the mouse (resembling what is done in AFM manipulation of
molecules).  The GUI can be run using the command:
\begin{verbatim}
  bash FireCore/tests/tMolGUIapp/run.sh
\end{verbatim}

Beside the above-mentioned functionalities, FireCore also implements
other advanced features like the charge equilibration
scheme\cite{Rappe91}, a workflow for fitting hydrogen bond corrections
from reference data (currently under development), the thermodynamic
integration method for the calculation of free energy differences, and
a QM/MM framework with an interface to the Fireball DFT(B)
package\cite{Lewis11, Mendieta14} and integrated high-resolution AFM
simulations.

\section{Conclusions and Outlook}\label{sec:outlook}

In this work, we have demonstrated that grid\-/projected force fields
-- originally inspired by methods used for ligand docking in molecular
biology -- can be successfully applied to surface science,
particularly for simulating molecular adsorption and manipulation on
surfaces using scanning probe microscopy at low temperatures.  As
implemented in the new open\-/source simulation package FireCore, the
method achieves a speedup of 2-3 orders of magnitude (compared to
conventional all\-/atom simulations) for a PTCDA molecule on a NaCl
surface using a single CPU, while maintaining a high accuracy of the
results.  We further showcase the CPU and GPU implementations of the
method in FireCore which enables sampling of millions of molecular
configurations per second, making it possible to exhaustively explore
all local minima of small flexible molecules (\emph{e.g.}, xylitol)
within just a few minutes on a standard desktop GPU.

The main limitation of the method is the assumption of a rigid
substrate. However, this is a common approximation even in traditional
all\-/atom simulations, particularly due to the lack of accurate force
fields for ionic crystal surfaces.  This limitation can be partially
mitigated by adjusting force field parameters to emulate the effective
(softened) potential resulting from substrate atom deflections -- a
strategy commonly employed in ligand docking\cite{Venkatachalam03}.
In a future work, we aim to address this limitation more rigorously by
incorporating additional force field components that allow for local
polarization and substrate deflection, based on linear response
theory\cite{Evans11, Tabacchi02, Harshan22}.  We are also developing a
hydrogen\-/bond correction scheme within the GridFF framework, as well
as density\-/derived potentials such as FDBM\cite{Ellner19}, which
promise to deliver accuracy far beyond traditional pairwise
interactions, like Coulomb, Morse, or Lennard\-/Jones, while retaining
unparalleled computational performance.  Although in this study we
opted for a conservative approach to GridFF interpolation using cubic
B\-/splines with very fine grid spacing, we recognize opportunities to
further optimize the memory footprint and potentially improve the
speed via a better cache locality. This could be achieved by
experimenting with alternative interpolation strategies, including
power\-/transformed interpolations\cite{Minh18}, different forms of
potential factorization, or non\-/uniform grid spacing
schemes\cite{Piegl97}.


\begin{acknowledgement}
  The authors thank Mithun Manikandan for the RESP calculations.
  This work was supported by the Czech Science Foundation, Project
  22-06008M, and co-funded by the European Union (Physics for Future
  -- Grant Agreement No. 101081515).
  A part of the computational resources was provided by the e-INFRA CZ
  project (ID:90254), supported by the Ministry of Education, Youth
  and Sports of the Czech Republic.
\end{acknowledgement}

%
%
%

\bibliography{sn-article}

\begin{mcitethebibliography}{83}
\bibitem[not()]{note_lammps}
{In LAMMPS this can be easily done using the \texttt{neigh\_modify exclude}
  command, while in the other codes there is generally no native support, even
  though workarounds exist.}\relax
\mciteBstWouldAddEndPunctfalse
\mciteSetBstMidEndSepPunct{\mcitedefaultmidpunct}
{}{\mcitedefaultseppunct}\relax
\EndOfBibitem
\end{mcitethebibliography}

\end{document}